\documentclass[12 pt,twoside,aps,prd,amsmath,amssymb,
tightenlines,showpacs,showkeys,eqsecnum]{revtex4}

\usepackage{graphicx}
\usepackage{mathptmx}
\usepackage{bm}
\newcommand {\ve}{\varepsilon}
\newcommand {\pr}{\partial}

\newcommand {\cD}{\cal D}
\newcommand {\bg}{\bar \gamma}
\newcommand {\G}{\Gamma}
\newcommand {\bp}{\bar \psi}
\newcommand {\p}{\psi}

\def\myfigure #1#2#3#4
{\begin{figure}[ht]\begin{center}
\includegraphics[width=#2 \textwidth]{#1.eps}
\parbox[t]{#4\textwidth}{\caption{#3}\label{#1}}
\end{center}\end{figure}}

\def \myfigures #1#2#3#4#5#6#7#8
{\begin{figure}[ht]
    \begin{center}
        \includegraphics[width=#2 \textwidth]{#1.eps}
        \hfill
        \includegraphics[width=#6 \textwidth]{#5.eps}
        \parbox[t]{#4\textwidth}{\caption {#3}\label{#1}}
        \hfill
        \parbox[t]{#8\textwidth}{\caption {#7}\label{#5}}
    \end{center}
\end{figure} }

\begin{document}
\baselineskip -24pt
\title{Nonlinear Spinor Fields in Bianchi type-I spacetime reexamined}
\author{Bijan Saha}
\affiliation{Laboratory of Information Technologies\\
Joint Institute for Nuclear Research\\
141980 Dubna, Moscow region, Russia} \email{bijan@jinr.ru}
\homepage{http://bijansaha.narod.ru}

\begin{abstract}

The specific behavior of spinor field in curve space-time with the
exception of FRW model almost always gives rise to non-trivial
non-diagonal components of the energy-momentum tensor. This
non-triviality of non-diagonal components of the energy-momentum
tensor imposes some severe restrictions either on the spinor field
or on the metric functions. In this paper within the scope of an
anisotropic Bianchi type-I Universe we study the role of spinor
field in the evolution of the Universe. It is found that there exist
two possibilities. In one scenario the initially anisotropic
Universe evolves into an isotropic one asymptotically, but in this
case the spinor field itself undergoes some severe restrictions. In
the second scenario the isotropization takes places almost at the
beginning of the process.

\end{abstract}

\keywords{Spinor field, dark energy, anisotropic cosmological
models, isotropization}

\pacs{98.80.Cq}

\maketitle

\bigskip

\section{Introduction}

According to the inflationary scenario, it is believed that a scalar
field known as inflaton is responsible for a rapid accelerated
expansion of the early Universe \cite{guth,ratra,olive}. For the
inflationary mechanism to work there must exist a weakly coupled
scalar field which is initially at a false vacuum which leads to the
inflation until the right vacuum value is obtained. This
inflationary model solves the problem of flatness, isotropy of
microwave background radiation and unwanted relics. Contrary to the
prediction of the standard cosmological models, recent observations
showed an accelerated mode of expansion of the present day Universe
\cite{riess,perlmutter}. Though the existence of an inflationary
scenario is not of much concern, the question of where the scalar
field comes from and why it undergoes such a peculiar phase
transition from false to right vacuum still remains unanswered. This
leads cosmologists to reconsider alternative possibilities.

As one of the way out many specialists considered spinor field as an
alternative source. Being related to almost all stable elementary
particles such as proton, electron and neutrino, spinor field,
especially Dirac spin-$1/2$ play a principal role at the microlevel.
However, in cosmology, the role of spinor field was generally
considered to be restricted. Only recently, after some remarkable
works  by different authors
\cite{henneaux,ochs,saha1997a,saha1997b,saha2001a,saha2004a,
saha2004b,saha2006c,saha2006e,saha2007,saha2006d,greene,ribas,souza,kremer},
showing the important role that spinor fields play on the evolution
of the Universe, the situation began to change. This change of
attitude is directly related to some fundamental questions of modern
cosmology: (i) problem of initial singularity; (ii) problem of
isotropization and (iii) late time acceleration of the Universe.

{\bf (i) Problem of initial singularity:} One of the problems of
modern cosmology is the presence of initial singularity, which means
the finiteness of time. The main purpose of introducing a nonlinear
term in the spinor field Lagrangian is to study the possibility of
the elimination of initial singularity. In a number of papers
\cite{saha1997a,saha1997b,saha2001a,saha2004a,saha2004b} it was
shown that the introduction of spinor field with a suitable
nonlinearity into the system indeed gives rise to singularity-free
models of the Universe.

{\bf  (ii) problem of isotropization:} Although the Universe seems
homogenous and isotropic at present, it does not necessarily mean
that it is also suitable for a description of the early stages of
the development  of the Universe and there are no observational data
guaranteeing the isotropy in the era prior to the recombination. In
fact, there are theoretical arguments that support the existence of
an anisotropic phase that approaches an isotropic one \cite{misner}
. The observations from Cosmic Background Explorer's differential
radiometer have detected and measured cosmic microwave background
anisotropies in different angular scales. These anisotropies are
supposed to hide in their fold the entire history of cosmic
evolution dating back to the recombination era and are being
considered as indicative of the geometry and the content of the
universe. More about cosmic microwave background anisotropy is
expected to be uncovered by the investigations of microwave
anisotropy probe. There is widespread consensus among the
cosmologists that cosmic microwave background anisotropies in small
angular scales have the key to the formation of discrete structure.
It was found that the introduction of nonlinear spinor field
accelerates the isotropization process of the initially anisotropic
Universe \cite{saha2001a,saha2004a,saha2006c}.

{\bf  (iii) late time acceleration of the Universe:} Some recent
experiments detected an accelerated mode of expansion of the
Universe \cite{riess,perlmutter}. Detection and further experimental
reconfirmation of current cosmic acceleration pose to cosmology a
fundamental task of identifying and revealing the cause of such
phenomenon. This fact can be reconciled with the theory if one
assumes that the Universe id mostly filled with so-called dark
energy. This form of matter (energy) is not observable in laboratory
and it does not interact with electromagnetic radiation. These facts
played decisive role in naming this object. In contrast to dark
matter, dark energy is uniformly distributed over the space, does
not intertwine under the influence of gravity in all scales and it
has a strong negative pressure of the order of energy density. Based
on these properties, cosmologists have suggested a number of dark
energy models those are able to explain the current accelerated
phase of expansion of the Universe. In this connection a series of
papers appeared recently in the literature, where a spinor field was
considered as an alternative model for dark energy \cite{ribas,
saha2006d,saha2006e,saha2007}.

It should be noted that most of the works mentioned above were
carried out within the scope of Bianchi type-I cosmological model.
Results obtained using a spinor field as a source of Bianchi type-I
cosmological field can be summed up as follows: A suitable choice of
spinor field nonlinearity\\

(i) {\it accelerates the isotropization process}
\cite{saha2001a,saha2004a,saha2006c};\\

(ii) {\it gives rise to a singularity-free Universe}
\cite{saha2001a,saha2004a,saha2004b,saha2006c};\\

(iii) {\it generates late time acceleration}
\cite{ribas,saha2006e,saha2007,saha2006d,souza}.

Given the role that spinor field can play in the evolution of the
Universe, question that naturally pops up is, if the spinor field
can redraw the picture of evolution caused by perfect fluid and dark
energy, is it possible to simulate perfect fluid and dark energy by
means of a spinor field? Affirmative answer to this question was
given in the a number of papers
\cite{krechet,saha2010a,saha2010b,saha2011,saha2012}. In those
papers spinor description of matter such as perfect fluid and dark
energy was given and the evolution of the Universe given by
different Bianchi models was thoroughly studied. In almost all the
papers the spinor field was considered to be time-dependent
functions and its energy-momentum tensor was given by the diagonal
elements only. Some latest study shows that due to the specific
connection with gravitational field the energy-momentum tensor of
the spinor field possesses non-trivial non-diagonal components as
well. In this paper we study the role of non-diagonal components of
the energy-momentum tensor of the spinor field in the evolution of
the Universe. To our knowledge such study was never done previously.
In section II we give the spinor field Lagrangian in details. In
section III the system of Einstein-Dirac equations is solved for BI
metric without engaging the non-diagonal components of
energy-momentum tensor as it was done in previous works of many
authors. In section IV we analyze the role of non-diagonal
components of energy-momentum tensor on the evolution of the
Universe.

\section{Spinor field Lagrangian}

For a spinor field $\p$, the symmetry between $\p$ and $\bp$ appears
to demand that one should choose the symmetrized Lagrangian
\cite{kibble}. Keeping this in mind we choose the spinor field
Lagrangian as \cite{saha2001a}:
\begin{equation}
L = \frac{\imath}{2} \biggl[\bp \gamma^{\mu} \nabla_{\mu} \psi-
\nabla_{\mu} \bar \psi \gamma^{\mu} \psi \biggr] - m_{\rm sp} \bp
\psi - F, \label{lspin}
\end{equation}
where the nonlinear term $F$ describes the self-interaction of a
spinor field and can be presented as some arbitrary functions of
invariants generated from the real bilinear forms of a spinor field.
Since $\psi$ and $\psi^{\star}$ (complex conjugate of $\psi$) have
four component each, one can construct $4\times 4 = 16$ independent
bilinear combinations. They are
\begin{subequations}
\label{bf}
\begin{eqnarray}
 S&=& \bar \psi \psi\qquad ({\rm scalar}),   \\
  P&=& \imath \bar \psi \gamma^5 \psi\qquad ({\rm pseudoscalar}), \\
 v^\mu &=& (\bar \psi \gamma^\mu \psi) \qquad ({\rm vector}),\\
 A^\mu &=&(\bar \psi \gamma^5 \gamma^\mu \psi)\qquad ({\rm pseudovector}), \\
T^{\mu\nu} &=&(\bar \psi \sigma^{\mu\nu} \psi)\qquad ({\rm
antisymmetric\,\,\, tensor}),
\end{eqnarray}
\end{subequations}
where $\sigma^{\mu\nu}\,=\,(\imath/2)[\gamma^\mu\gamma^\nu\,-\,
\gamma^\nu\gamma^\mu]$. Invariants, corresponding to the bilinear
forms, are
\begin{subequations}
\label{invariants}
\begin{eqnarray}
I &=& S^2, \\
J &=& P^2, \\
I_v &=& v_\mu\,v^\mu\,=\,(\bar \psi \gamma^\mu \psi)\,g_{\mu\nu}
(\bar \psi \gamma^\nu \psi),\\
I_A &=& A_\mu\,A^\mu\,=\,(\bar \psi \gamma^5 \gamma^\mu
\psi)\,g_{\mu\nu}
(\bar \psi \gamma^5 \gamma^\nu \psi), \\
I_T &=& T_{\mu\nu}\,T^{\mu\nu}\,=\,(\bar \psi \sigma^{\mu\nu}
\psi)\, g_{\mu\alpha}g_{\nu\beta}(\bar \psi \sigma^{\alpha\beta}
\psi).
\end{eqnarray}
\end{subequations}

According to the Fierz identity,  among the five invariants only $I$
and $J$ are independent as all others can be expressed by them: $I_v
= - I_A = I + J$ and $I_T = I - J.$ Therefore, we choose the
nonlinear term $F$ to be the function of $I$ and $J$ only, i.e., $F
= F(I, J)$, thus claiming that it describes the nonlinearity in its
most general form. Indeed, without losing generality we can choose
$F = F(K)$, with $K = \{I,\,J,\,I+J,\,I-J\}$. Here $\nabla_\mu$ is
the covariant derivative of spinor field:
\begin{equation}
\nabla_\mu \psi = \frac{\partial \psi}{\partial x^\mu} -\G_\mu \psi,
\quad \nabla_\mu \bp = \frac{\partial \bp}{\partial x^\mu} + \bp
\G_\mu, \label{covder}
\end{equation}
with $\G_\mu$ being the spinor affine connection. In \eqref{lspin}
$\gamma$'s are the Dirac matrices in curve space-time and obey the
following algebra
\begin{equation}
\gamma^\mu \gamma^\nu + \gamma^\nu \gamma^\mu = 2 g^{\mu\nu}
\label{al}
\end{equation}
and are connected with the flat space-time Dirac matrices $\bg$ in
the following way
\begin{equation}
 g_{\mu \nu} (x)= e_{\mu}^{a}(x) e_{\nu}^{b}(x) \eta_{ab},
\quad \gamma_\mu(x)= e_{\mu}^{a}(x) \bg_a, \label{dg}
\end{equation}
where $\eta_{ab}= {\rm diag}(1,-1,-1,-1)$ and $e_{\mu}^{a}$ is a set
of tetrad 4-vectors. The spinor affine connection matrices $\G_\mu
(x)$ are uniquely determined up to an additive multiple of the unit
matrix by the equation
\begin{equation}
\frac{\pr \gamma_\nu}{\pr x^\mu} - \G_{\nu\mu}^{\rho}\gamma_\rho -
\G_\mu \gamma_\nu + \gamma_\nu \G_\mu = 0, \label{afsp}
\end{equation}
with the solution
\begin{equation}
\Gamma_\mu = \frac{1}{4} \bg_{a} \gamma^\nu \partial_\mu e^{(a)}_\nu
- \frac{1}{4} \gamma_\rho \gamma^\nu \Gamma^{\rho}_{\mu\nu},
\label{sfc}
\end{equation}

Varying \eqref{lspin} with respect to $\bp (\psi)$ one finds the
spinor field equations:
\begin{subequations}
\label{speq}
\begin{eqnarray}
\imath\gamma^\mu \nabla_\mu \psi - m_{\rm sp} \psi - 2 F_K (S K_I +
 \imath P K_J \gamma^5) \psi &=&0, \label{speq1} \\
\imath \nabla_\mu \bp \gamma^\mu +  m_{\rm sp} \bp + 2 F_K \bp(S K_I
+  \imath P K_J \gamma^5) &=& 0. \label{speq2}
\end{eqnarray}
\end{subequations}
Here we denote $F_K = dF/dK$, $K_I = dK/dI$ and $K_J = dK/dJ.$

The energy-momentum tensor of the spinor field is given by
\begin{equation}
T_{\mu}^{\rho}=\frac{\imath}{4} g^{\rho\nu} \biggl(\bp \gamma_\mu
\nabla_\nu \psi + \bp \gamma_\nu \nabla_\mu \psi - \nabla_\mu \bar
\psi \gamma_\nu \psi - \nabla_\nu \bp \gamma_\mu \psi \biggr) \,-
\delta_{\mu}^{\rho} L \label{temsp}
\end{equation}
where $L$ in view of \eqref{speq} can be rewritten as
\begin{eqnarray}
L & = & \frac{\imath}{2} \bigl[\bp \gamma^{\mu} \nabla_{\mu} \psi-
\nabla_{\mu} \bar \psi \gamma^{\mu} \psi \bigr] - m_{\rm sp} \bp
\psi - F(K)
\nonumber \\
& = & \frac{\imath}{2} \bp [\gamma^{\mu} \nabla_{\mu} \psi - m_{\rm
sp} \psi] - \frac{\imath}{2}[\nabla_{\mu} \bar \psi \gamma^{\mu} +
m_{\rm sp} \bp] \psi
- F(K),\nonumber \\
& = & 2 F_K (I K_I + J K_J) - F = 2 K F_K - F(K). \label{lspin01}
\end{eqnarray}

We consider the case when the spinor field depends on $t$ only. In
this case for the components of energy-momentum tensor we find
\begin{subequations}
\begin{eqnarray}
T_0^0 &=& m_{\rm sp} S + F(K), \label{t00s}\\
T_1^1 = T_2^2 = T_3^3 &=&  F(K) - 2 K F_K. \label{t11s}
\end{eqnarray}
\end{subequations}

Let us now recall that in the unified nonlinear spinor theory of
Heisenberg, the massive term remains absent, and according to
Heisenberg, the particle mass should be obtained as a result of
quantization of spinor prematter~ \cite{massless,massless1}. In the
nonlinear generalization of classical field equations, the massive
term does not possess the significance that it possesses in the
linear one, as it by no means defines total energy (or mass) of the
nonlinear field system. Moreover, it was established that only a
massless spinor field with the Lagrangian \eqref{lspin} describes
perfect fluid from phantom to ekpyrotic matter
\cite{krechet,saha2010a,saha2010b,saha2011,saha2012}. Thus without
losing the generality we can consider the massless spinor field
putting $m\,=\,0.$

Inserting \eqref{t00s} and \eqref{t11s} into the barotropic equation
of state
\begin{equation}
p = W \ve, \label{beos}
\end{equation}
where $W$ is a constant, one finds
\begin{equation}
2 K F_K  = (1 + W) F(K), \label{beos1}
\end{equation}
with the solution
\begin{equation}
F(K) = \lambda K^{(1+W)/2}, \quad \lambda = {\rm const.}
\label{sol1}
\end{equation}
Depending on the value of $W$ \eqref{beos} describes perfect fluid
from phantom to ekpyrotic matter, namely
\begin{subequations}
\label{zeta}
\begin{eqnarray}
W &=& 0, \qquad \qquad \qquad {\rm (dust)},\\
W &=& 1/3, \quad \qquad \qquad{\rm (radiation)},\\
W &\in& (1/3,\,1), \quad \qquad\,\,{\rm (hard\,\,Universe)},\\
W &=& 1, \quad \qquad \quad \qquad {\rm (stiff \,\,matter)},\\
W &\in& (-1/3,\,-1), \quad \,\,\,\,{\rm (quintessence)},\\
W &=& -1, \quad \qquad \quad \quad{\rm (cosmological\,\, constant)},\\
W &<& -1, \quad \qquad \quad \quad{\rm (phantom\,\, matter)},\\
W &>& 1, \quad \qquad \quad \qquad{\rm (ekpyrotic\,\, matter)}.
\end{eqnarray}
\end{subequations}

In account of it the spinor field Lagrangian now reads
\begin{equation}
L = \frac{i}{2} \biggl[\bp \gamma^{\mu} \nabla_{\mu} \psi-
\nabla_{\mu} \bar \psi \gamma^{\mu} \psi \biggr] - \lambda
K^{(1+W)/2}. \label{lspin1}
\end{equation}
Thus a massless spinor field with the Lagrangian \eqref{lspin1}
describes perfect fluid from phantom to ekpyrotic matter. Here the
constant of integration $\lambda$ can be viewed as constant of
self-coupling. A detailed analysis of this study was given in
\cite{krechet,saha2010a,saha2010b,saha2011}.

A Chaplygin gas is usually described by a equation of state
\begin{equation}
p = -A/\ve^\gamma. \label{chap}
\end{equation}
Then in case of a massless spinor field for $F$ one finds
\begin{equation}
\frac{F^\gamma dF}{F^{1+\gamma} - A} = \frac{1}{2}\frac{dK}{K},
\label{eqq}
\end{equation}
with the solution \cite{saha2010a,saha2010b,saha2011}
\begin{equation}
F = \bigl(A + \lambda K^{(1+\gamma)/2}\bigr)^{1/(1+\gamma)}.
\label{chapsp}
\end{equation}
The Spinor field Lagrangian in this case takes the form
\begin{equation}
L = \frac{i}{2} \biggl[\bp \gamma^{\mu} \nabla_{\mu} \psi-
\nabla_{\mu} \bar \psi \gamma^{\mu} \psi \biggr] - \bigl(A + \lambda
K^{(1+\gamma)/2}\bigr)^{1/(1+\gamma)}. \label{lspin2}
\end{equation}

Finally, it should be noted that a quintessence with a modified
equation of state
\begin{equation}
p =  W (\ve - \ve_{\rm cr}), \quad W \in (-1,\,0), \label{mq}
\end{equation}
where $\ve_{\rm cr}$ some critical energy density, the spinor field
nonlinearity takes the form
\begin{equation}
F = \lambda S^{1+W} + \frac{W}{1+W}\ve_{\rm cr}. \label{Fmq}
\end{equation}
The spinor field Lagrangian in this case reads
\begin{equation}
L = \frac{i}{2} \biggl[\bp \gamma^{\mu} \nabla_{\mu} \psi-
\nabla_{\mu} \bar \psi \gamma^{\mu} \psi \biggr] - \lambda
K^{(1+W)/2} -\frac{W}{1+W}\ve_{\rm cr}. \label{lspin3}
\end{equation}
Setting $\ve_{\rm cr} = 0$ one gets \eqref{lspin1}. The purpose of
introducing the modified EoS was to avoid the problem of eternal
acceleration.

A detailed study of nonlinear spinor field was carried out in
\cite{saha2001a,saha2004a,saha2006c}. In what follows, exploiting
the equation of states we find the concrete form of $F$ which
describes various types of perfect fluid and dark energy.

\section{Bianchi type-I anisotropic cosmological model}

Let us study the evolution of the Universe filled with spinor field.
In doing so we consider the case when the gravitational field is
given by an anisotropic Bianchi type-I cosmological model.

Bianchi type-I (BI) model is the simplest anisotropic cosmological
model and gives an excellent scope to take into account the initial
anisotropy of the Universe. Given the importance of BI model to
study the effects of initial anisotropy in the evolution of he
Universe, we study this models in details.

We consider the BI metric in the form
\begin{equation}
ds^2 = dt^2 - a_1^{2} \,dx^{2} - a_2^{2} \,dy^{2} - a_3^{2}\,dz^2,
\label{bi}
\end{equation}
with $a_1\,a_2$ and $a_3$ being the functions of time only.

For further purpose we define the volume scale $V$ of the BI metric
as

\begin{equation}
V = abc. \label{Vdef}
\end{equation}

The system of Einstein equations in this case reads
\begin{subequations}
\label{BIE}
\begin{eqnarray}
\frac{\ddot a_2}{a_2} +\frac{\ddot a_3}{a_3} + \frac{\dot
a_2}{a_2}\frac{\dot a_3}{a_3}&=&  \kappa T_{1}^{1},\label{11bi}\\
\frac{\ddot a_3}{a_3} +\frac{\ddot a_1}{a_1} + \frac{\dot
a_3}{a_3}\frac{\dot a_1}{a_1}&=& \kappa T_{2}^{2},\label{22bi}\\
\frac{\ddot a_1}{a_1} +\frac{\ddot a_2}{a_2} + \frac{\dot
a_1}{a_1}\frac{\dot a_2}{a_2}&=&  \kappa T_{3}^{3},\label{33bi}\\
\frac{\dot a_1}{a_1}\frac{\dot a_2}{a_2} +\frac{\dot
a_2}{a_2}\frac{\dot a_3}{a_3} +\frac{\dot a_3}{a_3}\frac{\dot
a_1}{a_1}&=& \kappa T_{0}^{0}. \label{00bi}
\end{eqnarray}
\end{subequations}
Here $T_\mu^\nu$ is the energy momentum tensor of the spinor field.

Solving the Einstein equation on account of the fact that $T_1^1 =
T_2^2 = T_3^3$ for the metric functions one finds \cite{saha2001a}
\begin{eqnarray}
a_i = D_i V^{1/3} \exp{\Bigl(X_i \int \frac{dt}{V}\Bigr)}, \quad
\prod_{i=1}^{3} D_i = 1, \quad     \sum_{i=1}^{3} X_i = 0,
\label{metricf}
\end{eqnarray}
with $D_i$ and $X_i$ being the integration constants. Thus we se
that the metric functions can be expressed in terms of $V$.

Summation of \eqref{11bi}, \eqref{22bi}, \eqref{33bi} and 3 times
\eqref{00bi} leads to the equation for $V$  \cite{saha2001a}
\begin{eqnarray}
\ddot V = \frac{3 \kappa}{2} (T_0^0 + T_1^1) V. \label{detvbi}
\end{eqnarray}
As we have already found, the components of energy momentum tensor
are the function of $K$. If $K$ is a function of $V$, then the Eq.
\eqref{detvbi} possesses exact solution. In order to show that $K$
is a function of $V$ we go back to the spinor field equations. From
\eqref{speq} one dully finds
\begin{subequations}
\begin{eqnarray}
\dot S_0 - 4 F_K P K_J A_0 &=& 0, \label{S0}\\
\dot P_0 + 4 F_K S K_I A_0 &=& 0, \label{P0}\\
\dot A_0 - 4 F_K S K_I P_0 + 4 F_K P K_J S_0 &=& 0, \label{A0}
\end{eqnarray}
\end{subequations}
where $S_0 = S V$, $P_0 = P V$, $A_0 = A V$ with $A = \bp \gamma^0
\gamma^5 \psi$. Summation of \eqref{S0}, \eqref{P0} and \eqref{A0}
leads to
\begin{equation}
S^2 + P^2 + A^2 = C_1/V^2, \quad C_1 = {\rm const.} \label{SPA}
\end{equation}
On the other hand from \eqref{S0} and \eqref{P0} one finds
\begin{equation}
K_I S_0 \dot S_0 + K_J P_0 \dot P_0 = 0. \label{K0}
\end{equation}
In case of $K = I$, i.e., $K_I = 1$ and $K_J = 0$ from \eqref{K0}
one finds
\begin{equation}
K = I = S^2 = C_I/V^2, \quad C_I = {const.} \label{KI}
\end{equation}
For $K = J$, i.e., $K_I = 0$ and $K_J = 1$ from \eqref{K0} one finds
\begin{equation}
K = J = P^2 = C_J/V^2, \quad C_J = {const.} \label{KJ}
\end{equation}
If $K = I + J$, i.e., $K_I = 1$ and $K_J = 1$ from \eqref{K0} one
finds
\begin{equation}
K = I + J = S^2 + P^2 = C_{I+J}/V^2, \quad C_{I+J} = {const.}
\label{KIpJ}
\end{equation}
and finally, for $K = I - J$, i.e., $K_I = 1$ and $K_J = -1$ from
\eqref{K0} one finds
\begin{equation}
K = I - J = S^2 - P^2 = C_{I-J}/V^2, \quad C_{I-J} = {const.}
\label{KImJ}
\end{equation}
Thus we see that for the BI spacetime given by \eqref{bi} one finds
\begin{equation}
K = \frac{V_0^2}{V^2}, \quad V_0 = {\rm const.}  \label{KV}
\end{equation}

In case of \eqref{lspin1} we have
\begin{subequations}
\label{emtquint}
\begin{eqnarray}
T_0^0 &=& \varepsilon = \lambda K^{(1+W)/2}, \label{emtquint0}\\
T_1^1 &=& - p = - W \varepsilon = -W \lambda K^{(1+W)/2}.
\label{emtquint1}
\end{eqnarray}
\end{subequations}
Eq. \eqref{detvbi} then takes the form
\begin{equation}
\ddot V = \frac{3\kappa}{2}  \lambda V_0^{1+W} (1-W) V^{-W},
\label{Vquint}
\end{equation}
with the solution in quadrature
\begin{equation}
\int\frac{dV}{\sqrt{3 \kappa \lambda V_0^{1+W} V^{1-W} + C_1}} = t +
t_0. \label{Vquintquad}
\end{equation}
Here $C_1$ and $t_0$ are the integration constants.

Let us consider the case when the spinor field is given by the
Lagrangian \eqref{lspin2}. In this case we have
\begin{subequations}
\label{emtchap}
\begin{eqnarray}
T_0^0 &=& \varepsilon = \bigl(A + \lambda K^{(1+\gamma)/2}\bigr)^{1/(1+\gamma)}, \label{emtchap0}\\
T_1^1 &=& - p =  A/\varepsilon^\gamma =  A/\bigl(A + \lambda
K^{(1+\gamma)/2}\bigr)^{\gamma/(1+\gamma)}. \label{emtchap1}
\end{eqnarray}
\end{subequations}

The equation for $V$ now reads
\begin{equation}
\ddot V = \frac{3\kappa}{2} \Biggl[ \bigl(AV^{1+\gamma} + \lambda
V_0^{1+\gamma}\bigr)^{1/(1+\gamma)} + A
V^{1+\gamma}/\bigl(AV^{1+\gamma} + \lambda
V_0^{1+\gamma}\bigr)^{\gamma/(1+\gamma)}\Biggr], \label{Vchap}
\end{equation}
with the solution
\begin{equation}
\int \frac{dV}{\sqrt{C_1 + 3 \kappa V \bigl(AV^{1+\gamma} + \lambda
V_0^{1+\gamma}\bigr)^{1/(1+\gamma)}}} = t + t_0, \quad C_1 = {\rm
const}. \quad t_0 = {\rm const}. \label{Vchapquad}
\end{equation}
Inserting $\gamma = 1$ we come to the result obtained in
\cite{saha2005}.

Finally we consider the case with modified quintessence. Taking into
account that
\begin{subequations}
\begin{eqnarray}
T_0^0 &=& \lambda K^{(1+W)/2} + \frac{W}{1+W}\ve_{\rm cr}, \label{edmq}\\
T_1^1 = T_2^2 = T_3^3 &=& - \lambda  W K^{(1+W)/2} +
\frac{W}{1+W}\ve_{\rm cr}, \label{prmq}
\end{eqnarray}
\end{subequations}
for $V$ in this case we find
\begin{equation}
\ddot V = \frac{3\kappa}{2} \Bigl[\lambda V_0^{1 + W} (1 - W) V^{-W}
+ 2W \ve_{\rm cr}V/(1 + W)\Bigr], \label{Vmodq}
\end{equation}
with the solution in quadrature
\begin{equation}
\int \frac{dV}{\sqrt{3 \kappa \bigl[\lambda V_0^{1-W} V^{1 - W} +
W\ve_{\rm cr}V^2/(1 + W)\bigr]  + C_1}} = t + t_0. \label{qdmq}
\end{equation}
Here $C_1$ and $t_0$ are the integration constants. Comparing
\eqref{qdmq} with those with a negative $\Lambda$-term we see that
$\ve_{\rm cr}$ plays the role of a negative cosmological constant.

Let us also write the components of the spinor field explicitly. Let
us note that the spinor affine coefficients in case of BI metric
\eqref{bi} read
\begin{equation}
\G_0 = 0,\quad \G_1 = \frac{\dot a_1}{2} \bg^1 \bg^0, \quad \G_2 =
\frac{\dot a_2}{2} \bg^2 \bg^0, \quad \G_3 = \frac{\dot a_3}{2}
\bg^3 \bg^0. \label{sacbi}
\end{equation}

Then in view of \eqref{covder} and \eqref{sacbi} the spinor field
equation \eqref{speq1} takes the form
\begin{equation}
\imath \gamma^0 \bigl(\dot \psi + \frac{1}{2}\frac{\dot V}{V}
\psi\bigr) - m_{\rm sp} \psi - 2 F_K S K_I \psi - 2 \imath F_K P K_J
\gamma^5 \psi = 0. \label{speq1p}
\end{equation}
Further defining $\phi = \sqrt{V} \psi$ from \eqref{speq1p} one
finds
\begin{equation}
\imath \gamma^0 \dot \phi  - m_{\rm sp} \phi - 2 F_K S K_I \phi - 2
\imath F_K P K_J \gamma^5 \phi = 0. \label{speq1ph}
\end{equation}
For simplicity, we consider the case when $K = I$. As we have
already mentioned, $\psi$ is a function of $t$ only. We consider the
4-component spinor field given by
\begin{eqnarray}
\psi = \left(\begin{array}{c} \psi_1\\ \psi_2\\ \psi_3 \\
\psi_4\end{array}\right). \label{psi}
\end{eqnarray}
Taking into account that $\phi_i =\sqrt{V} \psi_i$ and defining
${\cD} = 2 S F_K K_I = 2 S F_K$ and inserting \eqref{psi} into
\eqref{speq1ph} in this case we find
\begin{subequations}
\label{speq1pfg}
\begin{eqnarray}
\dot \phi_1 + \imath {\cD} \phi_1 &=& 0, \label{ph1}\\
\dot \phi_2 + \imath {\cD} \phi_2 &=& 0, \label{ph2}\\
\dot \phi_3 - \imath {\cD} \phi_3 &=& 0, \label{ph3}\\
\dot \phi_4 - \imath {\cD} \phi_4 &=& 0, \label{ph4}.
\end{eqnarray}
\end{subequations}
Here we also consider the massless spinor field setting $m_{\rm sp}
= 0.$ The foregoing system of equations can be easily solved.
Finally for the spinor field we obtain
\begin{subequations}
\label{psinl}
\begin{eqnarray}
\psi_1(t) &=& (C_1/\sqrt{V}) \exp{\Biggl(-i\int {\cD} dt\Biggr)},\\
\psi_2(t) &=& (C_2/\sqrt{V}) \exp{\,\Biggl(-i\int {\cD} dt\Biggr)},\\
\psi_3(t) &=& (C_3/\sqrt{V}) \exp{\,\Biggl(i\int  {\cD} dt\Biggr)},\\
\psi_4(t) &=& (C_4/\sqrt{V}) \exp{\,\Biggl(i\int  {\cD} dt\Biggr)},
\end{eqnarray}
\end{subequations}
with $C_1,\,C_2,\,C_3,\,C_4$ being the integration constants and
related to $V_0$ as
$$C_1^* C_1 + C_2^* C_2 - C_3^* C_3 - C_4^* C_4 = V_0.$$
Thus we see that both the components of the spinor field as well as
the metric functions are the functions of $V$. It can be shown that
other physical quantities such as charge, spin current, spin and the
invariants of space-time are also the explicit function of $V$. It
was shown in previous papers that at any space-time point where $V =
0$ there occurs a space-time singularity \cite{saha2001a}. But in
all other cases ($V$ is the volume scale, hence should be
essentially non-negative), there exists unique solutions (for the
concrete values of problem parameters) to the equations for $V$,
i.e., \eqref{Vquint}, \eqref{Vchap}, and \eqref{Vmodq}, respectively
[cf Appendix B].

In what follows we will study the obtained results within the scope
of some recent findings, namely the fact that the spinor field
possesses noz-trivial non-diagonal components of the energy-momentum
tensor.

\section{What's wrong?}

In first view everything looks good and the papers written till the
date on this subject seems correct. But there is still something to
be worried about. In what follows, we speak about the new findings
on this field.

It should be remembered that the spinor field is more sensitive to
the gravitational one. It is due to specific spinor connection in
curve space-time. So, let us first write the spin affine connection
explicitly. For BI metric it looks:
\begin{equation}
\G_0 = 0, \quad \G_i = \frac{\dot a_i}{2} \bg^i \bg^0, \label{sac}
\end{equation}
Taking it into account from \eqref{temsp} it can be easily verified
that the energy-momentum tensor of the spinor field possesses
non-trivial non-diagonal components as well [cf. Appendix A].

\begin{subequations}
\begin{eqnarray}
T_0^0 & = & m_{\rm sp} S + F(K) \equiv F(K), \label{emt00}\\
T_1^1 &=& T_2^2 = T_3^3 = 2 K F_K - F(K), \label{emtii}\\
T_2^1 &=& \frac{\imath}{4} \frac{a_2}{a_1} \biggl(\frac{\dot
a_1}{a_1} - \frac{\dot a_2}{a_2}\biggr) \bp \bg^1 \bg^2 \bg^0 \psi, \label{emt12}\\
T_3^1 &=&\frac{\imath}{4} \frac{a_3}{a_1}
\biggl(\frac{\dot a_3}{a_3} - \frac{\dot a_1}{a_1}\biggr) \bp \bg^3 \bg^1 \bg^0 \psi, \label{emt13}\\
T_3^2 &=&\frac{\imath}{4} \frac{a_3}{a_2} \biggl(\frac{\dot
a_2}{a_2} - \frac{\dot a_3}{a_3}\biggr) \bp \bg^2 \bg^3 \bg^0 \psi.
\label{emt23}
\end{eqnarray}
\end{subequations}
So the complete set of Einstein equation for BI metric should be
\begin{subequations}
\label{BIEn}
\begin{eqnarray}
\frac{\ddot a_2}{a_2} +\frac{\ddot a_3}{a_3} + \frac{\dot
a_2}{a_2}\frac{\dot a_3}{a_3}&=&  \kappa (2 K F_K - F(K)),\label{11bin}\\
\frac{\ddot a_3}{a_3} +\frac{\ddot a_1}{a_1} + \frac{\dot
a_3}{a_3}\frac{\dot a_1}{a_1}&=& \kappa (2 K F_K - F(K)),\label{22bin}\\
\frac{\ddot a_1}{a_1} +\frac{\ddot a_2}{a_2} + \frac{\dot
a_1}{a_1}\frac{\dot a_2}{a_2}&=&  \kappa (2 K F_K - F(K)),\label{33bin}\\
\frac{\dot a_1}{a_1}\frac{\dot a_2}{a_2} +\frac{\dot
a_2}{a_2}\frac{\dot a_3}{a_3} +\frac{\dot a_3}{a_3}\frac{\dot
a_1}{a_1}&=& \kappa (m_{\rm sp} S + F(K)) \equiv \kappa F(K), \label{00bin}\\
0 &=& \frac{\imath}{4} \frac{a_2}{a_1} \biggl(\frac{\dot a_1}{a_1} -
\frac{\dot a_2}{a_2}\biggr) \bp \bg^1 \bg^2 \bg^0 \psi,
\label{emt12n}\\
0 &=&\frac{\imath}{4} \frac{a_3}{a_1} \biggl(\frac{\dot a_3}{a_3} -
\frac{\dot a_1}{a_1}\biggr) \bp \bg^3 \bg^1 \bg^0 \psi,
\label{emt13n}\\
0 &=&\frac{\imath}{4} \frac{a_3}{a_2} \biggl(\frac{\dot a_2}{a_2} -
\frac{\dot a_3}{a_3}\biggr) \bp \bg^2 \bg^3 \bg^0 \psi.
\label{emt23n}
\end{eqnarray}
\end{subequations}
In \eqref{00bin} we set the spinor mass $m_{\rm sp} = 0$. The
equations \eqref{emt12n}, \eqref{emt13n} and \eqref{emt23n} impose
some severe restrictions either on the spinor field, or on the
metric functions, or on both of them.

If the restrictions are imposed on the spinor field, we obtain
\begin{equation}
\bp \bg^1 \bg^2 \bg^0 \psi = \bp \bg^3 \bg^1 \bg^0 \psi = \bp \bg^2
\bg^3 \bg^0 \psi = 0. \label{spinor123}
\end{equation}
In this case the expressions obtained for the metric functions in
the earlier papers remain unaffected. But the components of the
spinor field will undergo some changes. It should be verified that
in this case the integration constants in \eqref{psinl} should obey
\begin{subequations}
\label{spcons}
\begin{eqnarray}
C_1^* C_1 - C_2^* C_2 - C_3^* C_3 + C_4^* C_4 &=& 0,\label{spcons1}\\
C_1^* C_2 - C_2^* C_1 + C_3^* C_4 - C_4^* C_3 &=& 0,
\label{spcons2}\\
C_1^* C_2 + C_2^* C_1 + C_3^* C_4 + C_4^* C_3 &=& 0, \label{spcons3}\\
C_1^* C_1 + C_2^* C_2 - C_3^* C_3 - C_4^* C_4 &=& V_0,
\label{spcons4}
\end{eqnarray}
\end{subequations}
which gives
\begin{subequations}
\label{spconsnew}
\begin{eqnarray}
C_1^* C_2 + C_3^* C_4 &=&  C_2^* C_1 + C_4^* C_3 = 0,
\label{spconsnew2}\\
C_1^* C_1 - C_3^* C_3 &=&  C_2^* C_2  - C_4^* C_4 = \frac{V_0}{2}.
\label{spconsnew4}
\end{eqnarray}
\end{subequations}

In the cases considered above for volume scale we obtained the
expressions given by \eqref{Vquintquad},\eqref{Vchapquad} and
\eqref{qdmq}, for the Universe filled with quintessence, Chaplygin
gas and quintessence with modified equation of state, respectively.
The Universe in these cases is initially anisotropic which evolves
into an isotropic one asymptotically \cite{saha2011,saha2012}.

The other possibility is to keep the components of the spinor field
unaltered. In this case from \eqref{emt12n}, \eqref{emt13n} and
\eqref{emt23n} for the metric functions one immediately finds:
\begin{equation}
\frac{\dot a_1}{a_1} = \frac{\dot a_2}{a_2} = \frac{\dot a_3}{a_3}
\equiv \frac{\dot a}{a}. \label{dota123}
\end{equation}
Taking into account that
$$\frac{\ddot a_i}{a_i} = \frac{d}{dt}\Bigl(\frac{\dot
a_i}{a_i}\Bigr) + \Bigl(\frac{\dot a_i}{a_i}\Bigr)^2 =
\frac{d}{dt}\Bigl(\frac{\dot a}{a}\Bigr) + \Bigl(\frac{\dot
a}{a}\Bigr)^2 = \frac{\ddot a}{a},$$

the system \eqref{BIEn} can be written as a system of two equations:
\begin{subequations}
\label{BIEn1}
\begin{eqnarray}
2\frac{\ddot a}{a} + \frac{\dot
a^2}{a^2}&=&  \kappa T_1^1,\label{11binf}\\
3\frac{\dot a^2}{a^2} &=& \kappa  T_0^0. \label{00binf}
\end{eqnarray}
\end{subequations}

In order to find the solution that satisfies both \eqref{11binf} and
\eqref{00binf} we rewrite \eqref{11binf} in view of \eqref{00binf}
in the following form:
\begin{equation}
\ddot a = \frac{\kappa}{6}\Bigl(3 T_1^1 - T_0^0\Bigr) a. \label{dda}
\end{equation}
Thus in account of non-diagonal components of the spinor field, we
though begin with Bianchi type-I space time, in reality solving the
Einstein field equations for FRW model. Before solving the equation
\eqref{dda}, let us go back to \eqref{metricf}. Taking into account
that
\begin{equation}
\frac{\dot a_i}{a_i} = \frac{\dot V}{3V} + \frac{X_i}{V},
\label{dotai}
\end{equation}
in view of \eqref{dota123} we find that
\begin{equation}
X_1 = X_2 = X_3 = 0. \label{Xi}
\end{equation}
The triviality of the integration constant $X_i$ follows from the
fact that $X_1 + X_2 + X_3 = 0$. Thus the solution \eqref{metricf}
should be written as
\begin{eqnarray}
a_i = D_i V^{1/3} = D_i a, \quad \prod_{i=1}^{3} D_i = 1,
\label{metricf1}
\end{eqnarray}
which means it represents a tiny sector of the general solutions
\eqref{metricf} which one obtains for the BI model in case of
isotropic distribution of matter with trivial non-diagonal
components of energy-momentum tensor, e.g., when the Universe is
filled with perfect fluid, dark energy etc.

Let us now define $a = V^{1/3}$ for different cases. In doing so we
recall that $K$ in this case takes the form
\begin{equation}
K = \frac{a_0^6}{a^6}, \quad a_0 = {\rm const.}. \label{Ka}
\end{equation}

Then then equation for $a$ in case of the spinor field given by
\eqref{lspin1} takes the form

\begin{equation}
\ddot a = - \frac{\kappa \lambda}{6} (1 + 3W) a_0^{3(1+W)}a^{-(2 + 3
W)}, \label{aquint}
\end{equation}
with the solution is quadrature
\begin{equation}
\int \frac{d a}{\sqrt{(\kappa \lambda/3) a_0^{3(1+W)}a^{-(1 + 3 W)}
+ E_1}} = t, \label{qudquint}
\end{equation}
with $E_1$ being integration constant.

As far as Chaplygin scenario is concerned in this case we have

\begin{equation}
\ddot a = - \frac{\kappa}{6} \frac{2A a^{3(1+\gamma)}- \lambda
a_0^{3(1+\gamma)}}{a^2 \bigl(A a^{3(1+\gamma)}- \lambda
a_0^{3(1+\gamma)}\bigr)^{\gamma/(1+\gamma)}}. \label{achap}
\end{equation}
This equation can be solved numerically.

Finally we consider the case with modified quintessence. Inserting
\eqref{edmq} and \eqref{prmq} into \eqref{dda} in this case we find
\begin{equation}
\ddot a  = -\frac{\kappa}{6}\Bigr[(3W+1)\lambda a_0^{3(1+W)}
a^{-(3W+2)} - \frac{2W}{1+W}\ve_{\rm cr} a\Bigr], \label{FRWmq}
\end{equation}
with he solution
\begin{equation}
\int\frac{d a}{\sqrt{(\kappa/3)\Bigl[\lambda a_0^{3(1+W)} a^{-(3W
+1)} + [W/(1+W)] \ve_{\rm cr} a^2 + E_2\Bigr]}} = t, \quad E_2 =
{\rm const}. \label{dda2}
\end{equation}
It can be shown that in case of modified quintessence the pressure
is sign alternating. As a result we have a cyclic mode of evolution.

\begin{figure}[ht]
\centering
\includegraphics[height=70mm]{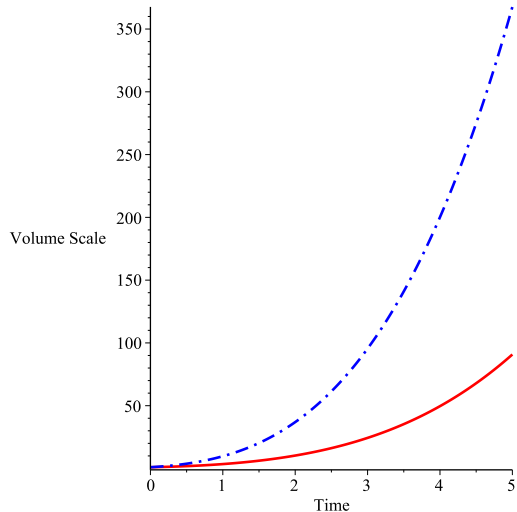} \\
\caption{Evolution of the Universe filled with quintessence. The
solid (red) line stands for volume scale $V$, while the dash-dot
(blue) line stands for $a^3$.} \label{quint2}.
\end{figure}

\begin{figure}[ht]
\centering
\includegraphics[height=70mm]{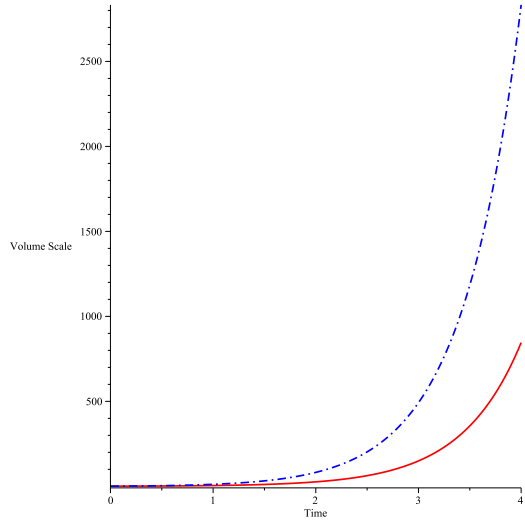} \\
\caption{Evolution of the Universe filled with Chaplygin gas. The
solid (red) line stands for volume scale $V$, while the dash-dot
(blue) line stands for $a^3$. } \label{chap2}.
\end{figure}

\begin{figure}[ht]
\centering
\includegraphics[height=70mm]{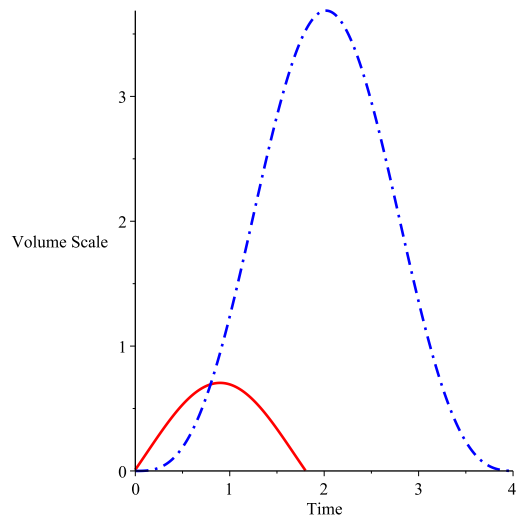} \\
\caption{Evolution of the Universe filled with quintessence with
modified equation of state. The solid (red) line stands for volume
scale $V$, while the dash-dot (blue) line stands for $a^3$. }
\label{mq2}.
\end{figure}

In what follows we illustrate the evolution of the Universe filled
with quintessence, Chaplygin gas and quintessence with modified
equation of state for two different cases: when the restrictions are
imposed on the spinor field and when the metric functions were
restricted. In Figures \ref{quint2}, \ref{chap2} and \ref{mq2} we
illustrated the evolution of the Universe filled with quintessence,
Chaplygin gas and quintessence with modified equation of state,
respectively. The solid (red) line stands for the volume scale, when
the restrictions due to non-zero non-diagonal components of the
energy momentum tensor of the spinor field, were imposed on the
components of the spinor field. In this case the isotropization
takes place asymptotically. The blue line shows the evolution of the
Universe when due to the non-zero non-diagonal components of the
energy momentum tensor of the spinor field leads to the immediate
isotropization of the Universe. Here we plot the volume scale as
$a^3$, which $a$ being the average scale factor.

As one sees, in case of early isotropization the Universe grows
rapidly.

\section{Conclusion}

Within the scope of Bianchi type-I space time we study the role of
spinor field on the evolution of the Universe. It is shown that even
in case of space independent of the spinor field it still possesses
non-zero non-diagonal components of energy-momentum tensor thanks to
its specific relation with gravitational field. This fact plays
vital role on the evolution of the Universe. There might be two
different scenarios. In one case only the components of the spinor
field are affected leaving the space-time initially anisotropic that
evolves into an isotropic one asymptotically. According to the
second scenario, the space-time becomes isotropic right from the
beginning,i.e.,
\begin{equation}
a_1 \sim a_2 \sim a_3, \label{a123sim}
\end{equation}
and can be completely described by the Einstein field equations for
FRW metric. As numerical analysis shows, in case of early
isotropization the Universe expands rather rapidly. There might be
another possibility when the non-diagonal components of
energy-momentum tensor influence both the spinor field and metric
functions simultaneously. Finally, it should be emphasized that the
spinor field Lagrangian \eqref{lspin} can be used to simulate a time
varying EoS parameter and DP as well. Models with time varying EoS
parameter and DP have been extensively studied in recent time
\cite{YRRG,YPS,YS,SY}. We plan to study all these possibilities
within the scope of different Bianchi models in our forthcoming
papers.

\vskip 0.1 cm

\noindent {\bf Acknowledgments}\\
This work is supported in part by a joint Romanian-LIT, JINR, Dubna
Research Project, theme no. 05-6-1060-2005/2013.

\section{appendix A}

\renewcommand{\theequation}{A.\arabic{equation}}

Since the energy-momentum tensor of the spinor field is not widely
discussed in literature, we consider it here in details. The
energy-momentum tensor of the spinor field is given by
\eqref{temsp}. Let us rewrite the expression once again
\begin{equation}
T_{\mu}^{\rho}=\frac{\imath}{4} g^{\rho\nu} \biggl(\bp \gamma_\mu
\nabla_\nu \psi + \bp \gamma_\nu \nabla_\mu \psi - \nabla_\mu \bar
\psi \gamma_\nu \psi - \nabla_\nu \bp \gamma_\mu \psi \biggr) \,-
\delta_{\mu}^{\rho} L \label{temsp1}
\end{equation}
In view of \eqref{covder}, i.e.,
\begin{equation}
\nabla_\mu \psi = \frac{\partial \psi}{\partial x^\mu} -\G_\mu \psi,
\quad \nabla_\mu \bp = \frac{\partial \bp}{\partial x^\mu} + \bp
\G_\mu, \label{covder1}
\end{equation}
\eqref{temsp1} can be rewritten as

\begin{eqnarray}
T_{\mu}^{\,\,\,\rho}&=&\frac{\imath}{4} g^{\rho\nu} \biggl(\bp
\gamma_\mu
\partial_\nu \psi + \bp \gamma_\nu \partial_\mu \psi -
\partial_\mu \bar \psi \gamma_\nu \psi - \partial_\nu \bp \gamma_\mu
\psi \biggr)\nonumber\\
& - &\frac{\imath}{4} g^{\rho\nu} \bp \biggl(\gamma_\mu \G_\nu +
\G_\nu \gamma_\mu + \gamma_\nu \G_\mu + \G_\mu \gamma_\nu\biggr)\psi
 \,- \delta_{\mu}^{\rho} L, \label{temsp0}\\
 &=&  \frac{\imath}{4} g^{\rho\nu} {\bar T}_{\nu\mu} - \frac{\imath}{4} g^{\rho\nu}
 {\tilde T}_{\nu\mu} - \delta_{\mu}^{\rho} L. \nonumber
\end{eqnarray}
Now for BI metric we have
\begin{equation}
\G_0 = 0, \quad \G_1 = \frac{\dot a_1}{2} \bg^1 \bg^0, \quad \G_2 =
\frac{\dot a_2}{2} \bg^2 \bg^0, \quad \G_3 = \frac{\dot a_3}{2}
\bg^3 \bg^0, \label{sac123}
\end{equation}

Tetrads are connected to the metric functions as
\begin{equation}
g_{\mu\nu} = e^{(a)}_\mu e^{(b)}_\nu \eta_{ab}, \label{tg}
\end{equation}
with $\eta_{ab} = {\rm diag}\,[1,\,-1,\,-1,\,-1]$ or $\eta_{ab} =
{\rm diag}\,[-1,\,1,\,1,\,1]$. Dirac matrices in flat space-time
$\bg_a$ are connected to those of in curved space-time as follows:
\begin{equation}
\gamma_\mu = e^{(a)}_\mu \bg_a. \label{gammabg}
\end{equation}
Beside these $\gamma_\mu$ and $\bg$ satisfy the following relations:
\begin{equation}
\gamma_\mu \gamma_\nu + \gamma_\nu\gamma_\mu = 2 g_{\mu \nu},
\label{ggam}
\end{equation}

\begin{equation}
\bg_a \bg_b + \bg_b \bg_a = 2 \eta_{ab}. \label{bget}
\end{equation}
We use $g_{\mu\nu}$ or $g^{\mu\nu}$ to lower to raise the indices of
$\gamma$ matrices, and  $\eta_{ab}$ or $\eta^{ab}$ to lower to raise
the indices of $\bg$ matrices:
\begin{eqnarray}
\gamma^\mu &=& g^{\mu\nu} \gamma_\nu, \qquad \gamma_\mu = g_{\mu\nu}
\gamma^\nu, \label{gud}\\
\bg^a &=& \eta^{ab} \bg_b, \qquad \bg_a = \eta_{ab} \bg^b.
\label{bgud}
\end{eqnarray}

In view of \eqref{tg} we choose the tetrad as follows:

\begin{equation}
e_0^{(0)} = 1, \quad e_1^{(1)} = a_1, \quad e_2^{(2)} = a_2, \quad
e_3^{(3)} = a_3. \label{tetradvi}
\end{equation}

From \eqref{gammabg} one now finds
\begin{equation}
\gamma_0 = \bg_0, \quad  \gamma_1 = a_1  \bg_1, \quad \gamma_2 = a_2
 \bg_2, \quad \gamma_3 = a_3 \bg_3. \label{gbgvi}
\end{equation}
Taking into account that in our case
$$\bg^0 = \bg_0, \quad \bg^1 = -\bg_1, \quad \bg^2 = -\bg_2, \quad
\bg^3 = -\bg_3,$$ one also finds
\begin{equation}
\gamma^0 = \bg^0, \quad  \gamma^1 = \frac{1}{a_1} \bg^1, \quad
\gamma^2 = \frac{1}{a_2} \bg^2, \quad \gamma^3 = \frac{1}{a_3}
\bg^3. \label{gbgviup}
\end{equation}

Finally, taking into account that $\bg^i \bg^i \bg^j + \bg^i \bg^j
\bg^i = 0$ and $\bg^i \bg^j \bg^k + \bg^j \bg^k \bg^i = 2 \bg^i
\bg^j \bg^k$, for $i \ne j \ne k = 0,\,1,\,2,\,3$ one finds
\begin{eqnarray}
\gamma_0 \G_0 + \G_0 \gamma_0 & = & 0, \nonumber\\
\gamma_1 \G_1 + \G_1 \gamma_1 & = & 0, \nonumber\\
\gamma_2 \G_2 + \G_2 \gamma_2 & = & 0, \nonumber\\
\gamma_3 \G_3 + \G_3 \gamma_3 & = & 0, \nonumber\\
\gamma_1 \G_2 + \G_2 \gamma_1 & = & -a_1 \dot a_2 \bg^1 \bg^2 \bg^0,
\nonumber\\
\gamma_2 \G_1 + \G_2 \gamma_1 & = & a_2 \dot a_1 \bg^1 \bg^2
\bg^0, \nonumber\\
\gamma_1 \G_3 + \G_3 \gamma_1 & = & a_1 \dot a_3 \bg^3 \bg^1 \bg^0, \nonumber\\
\gamma_3 \G_1 + \G_1 \gamma_3 & = & - a_3 \dot a_1 \bg^3 \bg^1 \bg^0, \nonumber\\
\gamma_2 \G_3 + \G_3 \gamma_2 & = & - a_2 \dot a_3 \bg^2 \bg^3 \bg^0, \nonumber\\
\gamma_3 \G_2 + \G_2 \gamma_3 & = &  a_3 \dot a_2 \bg^2 \bg^3 \bg^0, \nonumber\\
\gamma_0 \G_1 + \G_1 \gamma_0 & = & 0, \nonumber\\
\gamma_0 \G_2 + \G_2 \gamma_0 & = & 0, \nonumber\\
\gamma_0 \G_3 + \G_3 \gamma_0 & = & 0. \nonumber
\end{eqnarray}
Hence we get 
\begin{subequations}
\label{tidT}
\begin{eqnarray}
 {\tilde T}_{00} & = &  2\bp \bigl(\gamma_0 \G_0 +
\G_0 \gamma_0\bigr)\psi = 0,\label{tidT00}\\
{\tilde T}_{11} & = &  2\bp \bigl(\gamma_1 \G_1 +
\G_1 \gamma_1\bigr)\psi = 0,\label{tidT11}\\
{\tilde T}_{22} & = &  2\bp \bigl(\gamma_2 \G_2 +
\G_2 \gamma_2\bigr)\psi = 0,\label{tidT22}\\
{\tilde T}_{33} & = &  2\bp \bigl(\gamma_3 \G_3 +
\G_3 \gamma_3\bigr)\psi = 0,\label{tidT33}\\
{\tilde T}_{03} &=& \bp \bigl(\gamma_0 \G_3 +
\G_3 \gamma_0 + \gamma_3 \G_0 + \G_0 \gamma_3\bigr)\psi = 0, \label{tidT03}\\
  {\tilde T}_{01} & = & \bp \bigl(\gamma_0 \G_1 +
\G_1 \gamma_0 + \gamma_1 \G_0 + \G_0 \gamma_1\bigr)\psi = 0, \label{tidT01}\\
 {\tilde T}_{02} & = & \bp \bigl(\gamma_0 \G_2 +
\G_2 \gamma_0 + \gamma_2 \G_0 + \G_0 \gamma_2\bigr)\psi = 0, \label{tidT02}\\
 {\tilde T}_{12} & = & \bp \bigl(\gamma_1 \G_2 +
\G_2 \gamma_1 + \gamma_2 \G_1 + \G_1 \gamma_2\bigr)\psi =
  a_1 a_2 \bigl(\frac{\dot a_1}{a_1} - \frac{\dot a_2}{a_2}\bigr)
 \bp \bg^1 \bg^2 \bg^0 \psi, \label{tidT12}\\
{\tilde T}_{23} & = & \bp \bigl(\gamma_3 \G_2 + \G_2 \gamma_3 +
\gamma_2 \G_3 + \G_3 \gamma_2\bigr)\psi =  a_2 a_3 \bigl(\frac{\dot
a_2}{a_2} - \frac{\dot a_3}{a_3}\bigr) \bp
\bg^2 \bg^3 \bg^0 \psi, \label{tidT23}\\
{\tilde T}_{31} & = &  \bp \bigl(\gamma_1 \G_3 + \G_3 \gamma_1 +
\gamma_3 \G_1 + \G_1 \gamma_3\bigr)\psi =  a_3 a_1\bigl(\frac{\dot
a_3}{a_3} - \frac{\dot a_1}{a_1}\bigr) \bp \bg^3 \bg^1 \bg^0 \psi.
\label{tidT31}
\end{eqnarray}
\end{subequations}

Let us now calculate ${\bar T}_{\nu\mu} = \bigl(\bp \gamma_\mu
\partial_\nu \psi + \bp \gamma_\nu \partial_\mu \psi -
\partial_\mu \bar \psi \gamma_\nu \psi - \partial_\nu \bp \gamma_\mu
\psi \bigr)$. Since $\psi$ is a function of $t$ only, i.e. $\psi
\psi (t)$ we immediately get ${\bar T}_{11} = {\bar T}_{22} = {\bar
T}_{33} = {\bar T}_{12}= {\bar T}_{23} = {\bar T}_{31} = 0$. Whereas
for the remaining components we obtain

\begin{eqnarray}
{\bar T}_{00} &=& 2 \bigl(\bp \gamma_0 \dot \psi - \dot \bp \gamma_0
\psi \bigr), \nonumber\\
{\bar T}_{01} &=&  \bigl(\bp \gamma_1 \dot \psi - \dot \bp \gamma_1
\psi \bigr), \nonumber\\
{\bar T}_{02} &=&  \bigl(\bp \gamma_2 \dot \psi - \dot \bp \gamma_2
\psi \bigr), \nonumber\\
{\bar T}_{03} &=&  \bigl(\bp \gamma_3 \dot \psi - \dot \bp \gamma_3
\psi \bigr), \nonumber
\end{eqnarray}

To estimate the foregoing quantities let us go back to the spinor
field equations \eqref{speq}, which we rewrite as
\begin{subequations}
\label{speqn1}
\begin{eqnarray}
\imath\gamma^0 \dot \psi + \frac{\imath}{2}\frac{\dot V}{V} \gamma^0
\psi - m_{\rm sp} \psi - 2 F_K (S K_I +
 \imath P K_J \gamma^5) \psi &=&0, \label{speq1n} \\
\imath \dot \bp \gamma^0 +  \frac{\imath}{2}\frac{\dot V}{V} \bp
\gamma^0  + m_{\rm sp} \bp + 2 F_K \bp(S K_I + \imath P K_J
\gamma^5) &=& 0, \label{speq2n}
\end{eqnarray}
\end{subequations}
where $V = a_1 a_2 a_3$. Multiplying \eqref{speq1n} by $\bp
\gamma^0$ from the left and \eqref{speq2n} by $\gamma^0 \psi$ from
the right and subtracting the second equation from the first we
obtain
\begin{equation}
\imath \bigl(\bp \gamma_0 \dot \psi - \dot \bp \gamma_0 \psi \bigr)
= 2 m_{\rm sp} S + 4 F_K (I K_I + J K_J) = 2 m_{\rm sp} S + 4 K F_K.
\label{00term}
\end{equation}
Multiplying \eqref{speq1n} by $\bp \gamma^j \gamma^0$ from the left
and \eqref{speq2n} by $\gamma^0 \gamma^j \psi$ from the right, where
$j = 1,2,3$ and subtracting the second equation from the first we
obtain
\begin{subequations}
\label{speqn}
\begin{eqnarray}
\imath \bigl(\bp \gamma_1 \dot \psi - \dot \bp \gamma_1 \psi \bigr)
= \bigl( m_{\rm sp} + 2 F_K S K_I\bigr) \bp (\gamma^1 \gamma^0 +
\gamma^0 \gamma^1) \psi + 2 \imath P K_J \bp (\gamma^1 \gamma^0
\gamma^5 + \gamma^5 \gamma^0 \gamma^1)\psi  &=&  0, \nonumber\\
\imath \bigl(\bp \gamma_2 \dot \psi - \dot \bp \gamma_2 \psi \bigr)
= \bigl( m_{\rm sp} + 2 F_K S K_I\bigr) \bp (\gamma^2 \gamma^0 +
\gamma^0 \gamma^2) \psi + 2 \imath P K_J \bp (\gamma^2 \gamma^0
\gamma^5 + \gamma^5 \gamma^0 \gamma^2)\psi  &=&  0, \nonumber\\
\imath \bigl(\bp \gamma_3 \dot \psi - \dot \bp \gamma_3 \psi \bigr)
= \bigl( m_{\rm sp} + 2 F_K S K_I\bigr) \bp (\gamma^3 \gamma^0 +
\gamma^0 \gamma^3) \psi + 2 \imath P K_J \bp (\gamma^3 \gamma^0
\gamma^5 + \gamma^5 \gamma^0 \gamma^3)\psi  &=&  0. \nonumber
\end{eqnarray}
\end{subequations}
Hence we get
\begin{eqnarray}
{\bar T}_{00} = -\imath(4 m_{\rm sp} S + 8 K F_K), \quad  {\bar
T}_{01} = 0, \quad {\bar T}_{02} = 0, \quad {\bar T}_{03} = 0.
\nonumber
\end{eqnarray}
Taking into account that $L = 2 K F_K - F(K)$ from \eqref{temsp0} we
find the following expressions for the components of the energy
momentum tensor:

\begin{subequations}
\label{Ttot}
\begin{eqnarray}
 T_0^0 & = &  \frac{\imath}{4} g^{00} {\bar T}_{00} - L  = m_{\rm sp} S + F(K),\label{Ttot00}\\
T_1^1 & = &  - L  = F(K) - 2 K F_K,\label{Ttot11}\\
T_2^2 & = &  -L = F(K) - 2 K F_K,\label{Ttot22}\\
T_3^3 & = &  -L = F(K) - 2 K F_K,\label{Ttot33}\\
T_3^0 &=& 0, \label{Ttot03}\\
  T_1^0 & = & 0, \label{Ttot01}\\
 T_2^0 & = &  0, \label{Ttot02}\\
 T_2^1 & = & - \frac{\imath}{4} g^{11}
 T_{12} = -\frac{\imath}{4} \frac{a_2}{a_1} \bigl(\frac{\dot a_1}{a_1} - \frac{\dot a_2}{a_2}\bigr)
 \bp \bg^1 \bg^2 \bg^0 \psi, \label{Ttot12}\\
T_3^2 & = & - \frac{\imath}{4} g^{22}
 T_{23} = -\frac{\imath}{4} \frac{a_3}{a_2} \bigl(\frac{\dot a_2}{a_2} - \frac{\dot a_3}{a_3}\bigr)
 \bp \bg^2 \bg^3 \bg^0 \psi, \label{Ttot23}\\
T_3^1 & = &  - \frac{\imath}{4} g^{11}
 T_{13} = -\frac{\imath}{4} \frac{a_3}{a_1} \bigl(\frac{\dot a_3}{a_3} - \frac{\dot a_1}{a_1}\bigr)
 \bp \bg^3 \bg^1 \bg^0 \psi. \label{Ttot31}
\end{eqnarray}
\end{subequations}

\section{appendix B}

\renewcommand{\theequation}{B.\arabic{equation}}

As it was shown earlier, the metric functions, components of the
spinor field, as well as other physical quantities such as charge,
spin current, spin, invariants of BI space-time explicitly depend on
$V$. Moreover, these quantities becomes zero at any space-time point
where $V = 0$, thus giving rise to a space-time singularity. So it
is important to study the equation for $V$  i.e., \eqref{Vquint},
\eqref{Vchap}, and \eqref{Vmodq} in details.

Let us prove the existence and uniqueness of the solution to the
equations for $V$. Since $V$ is the volume scale, it is essentially
non-negative. Taking into account that $V=0$ gives rise to a
space-time singularity, we consider the case when $V$ is positive.
Note that we are modeling an expanding or cyclic Universe and we can
choose the problem parameters in such a way that the equations for
$V$  i.e., \eqref{Vquint}, \eqref{Vchap}, and \eqref{Vmodq} allows
only positive $V$. For simplicity we consider the case with
quintessence \eqref{Vquint} and rewrite it in the form
\begin{equation}
\ddot V = A V^{-W}. \label{Vquint0}
\end{equation}
We show it using Lipshitz condition. The equation \eqref{Vquint0} we
rewrite in the form
\begin{subequations}
\label{Lip}
\begin{eqnarray}
\dot f &=& A V^{-W}, \label{Lip1}\\
\dot V &=& f, \label{Lip2}
\end{eqnarray}
\end{subequations}
or equivalently,
\begin{equation}
\left(\begin{array}{c} f\\ V \end{array}\right)^{\cdot} = F
\left(\begin{array}{c} V\\ f \end{array}\right) =
\left(\begin{array}{c} A V^{-W}\\ f \end{array}\right). \label{Lip0}
\end{equation}
According to Lipshitz condition there should exist a constant $M$
such that
\begin{equation}
\biggl| F \left(\begin{array}{c} V_{1} \\ f_{1} \end{array}\right) -
F \left(\begin{array}{c} V_{2} \\ f_{2} \end{array}\right)\biggr| <
M \biggl| \left(\begin{array}{c} V_{1}\\ f_{1} \end{array}\right) -
\left(\begin{array}{c} V_{2}\\ f_{2} \end{array}\right)\biggr|.
\label{LipCon}
\end{equation}
Using the mean value theorem we find
\begin{eqnarray}
\biggl| F \left(\begin{array}{c} V_{1} \\ f_{1} \end{array}\right) -
F \left(\begin{array}{c} V_{2} \\ f_{2} \end{array}\right)\biggr|
&=& \sqrt{\bigl(AV_1^{-W} - A V_2^{-W}\bigr)^2 + \bigl(f_1 -
f_2\bigr)^2} \nonumber\\ &=& \sqrt{A^2 \bigl(V_1 - V_2\bigr)^2
\bigl(-W V_{*}^{-W-1}\bigr)^2 + \bigl(f_1 - f_2\bigr)^2}, \label{rh}
\end{eqnarray}
where $V_{*}$ is some value of $V$ in between $V_1$ and $V_2$.
Inserting \eqref{rh} into \eqref{LipCon} we find
\begin{eqnarray}
A^2 \bigl(V_1 - V_2\bigr)^2 \bigl(-W V_{*}^{-W-1}\bigr)^2 +
\bigl(f_1 - f_2\bigr)^2 < M^2\biggl[\bigl(V_1 - V_2\bigr)^2 -
\bigl(f_1 - f_2\bigr)^2\biggr], \label{LC1}
\end{eqnarray}
from which follows
\begin{eqnarray}
A^2 \bigl(V_1 - V_2\bigr)^2 \bigl(-W V_{*}^{-W-1}\bigr)^2  <
M^2\biggl[\bigl(V_1 - V_2\bigr)^2 - (M^2 -1) \bigl(f_1 -
f_2\bigr)^2\biggr]. \label{LC2}
\end{eqnarray}
For the \eqref{LC2} holds, it is sufficient for $|M| > 1$ the
following relation:
\begin{eqnarray}
A^2 \bigl(V_1 - V_2\bigr)^2 \bigl(-W V_{*}^{-W-1}\bigr)^2  <
M^2\bigl(V_1 - V_2\bigr)^2, \label{LC3}
\end{eqnarray}
which leads to
\begin{eqnarray}
A^2 \bigl(-W V_{*}^{-W-1}\bigr)^2  < M^2. \label{LC4}
\end{eqnarray}
Hence we conclude that it is sufficient to take
\begin{equation}
|M| > \begin{array}{c} max \\ V_{*} \end{array} | A W V_{*}^{-W
-1}|. \label{max}
\end{equation}
Such a $M$ exists in the interval $[\epsilon,\,R]$, where $0 <
\epsilon < R$. Hence we conclude that there exists the unique
solution to the equation \eqref{Vquint0}.

\end{document}